\documentclass{article}

\usepackage{arxiv}

\usepackage[utf8]{inputenc} 
\usepackage[T1]{fontenc}    
\usepackage{hyperref}       
\usepackage{url}            
\usepackage{booktabs}       
\usepackage{amsfonts}       
\usepackage{nicefrac}       
\usepackage{microtype}      
\usepackage{lipsum}
\usepackage{graphicx}
\usepackage{amsmath}
\usepackage{breqn}
\graphicspath{ {./images/} }

\title{SiameseLSRTM: Enhancing least-squares reverse time migration with a Siamese network}

\author{
 Xinru Mu \\
  Physical Science and Engineering Division\\
  King Abdullah University of Science and Technology\\
  Thuwal 23955, Saudi Arabia \\
  \texttt{xinru.mu@kaust.edu.sa} \\
   \And
Omar M. Saad \\
  Physical Science and Engineering Division\\
  King Abdullah University of Science and Technology\\
  Thuwal 23955, Saudi Arabia \\
  \And
 Tariq Alkhalifah \\
  Physical Science and Engineering Division\\
  King Abdullah University of Science and Technology\\
  Thuwal 23955, Saudi Arabia \\
}

\begin{document}
\maketitle

\begin{abstract}
Least-squares reverse time migration (LSRTM) is an inversion-based imaging method rooted in optimization theory, which iteratively updates the reflectivity model to minimize the difference between observed and simulated data. However, in real data applications, the Born-based simulated data, based on simplified physics, like the acoustic assumption, often under represent the complexity within observed data. Thus, we develop SiameseLSRTM, a novel approach that employs a Siamese network consisting of two identical convolutional neural networks (CNNs) with shared weights to measure the difference between simulated and observed data. Specifically, the shared-weight CNNs in the Siamese network enable the extraction of comparable features from both observed and simulated data, facilitating more effective data matching and ultimately improving imaging accuracy. 
SiameseLSRTM is a self-supervised framework in which the network parameters are updated during the iterative LSRTM process, without requiring extensive labeled data and prolonged training. We evaluate SiameseLSRTM using two synthetic datasets and one field dataset from a land seismic survey, showing that it produces higher-resolution and more accurate imaging results compared to traditional LSRTM.
\end{abstract}


\section{Introduction}
Reverse time migration (RTM) plays a crucial role in exploration geophysics. By utilizing the two-way wave equation, it accurately images complex subsurface structures with steep dips, outperforming ray-based and one-way wave equation-based imaging methods \cite{baysal1983reverse, bednar2005brief, mcmechan2008migration, zhou2018reverse}. However, the migration operator in RTM is essentially the adjoint of Born-based modeling, which results in blurred imaging \cite{claerbout2014geophysical}. To improve imaging resolution, applying the inverse of the Hessian to the RTM image can significantly improve imaging accuracy \cite{tang2009target, schuster2017seismic}. Nevertheless, directly computing the inverse of the Hessian is computationally expensive and memory-intensive \cite{tang2009target}. As a more practical alternative, least-squares reverse time migration (LSRTM) addresses the high memory requirement by approximating the inverse of the Hessian via iterative processes of migration and demigration \cite{tarantola1984linearized, dai2011least, zhao2018fast, gu2017elastic, yang2019least}. LSRTM can suppress migration noise, balance amplitude, and enhance resolution, ultimately achieving high-accuracy imaging, making it widely adopted in industrial applications \cite{wong2011least, fletcher2016least, mu2020least, yang2020elastic, duveneck2021reflection}.

LSRTM is an optimization inversion method that iteratively updates the reflectivity model to minimize the difference between the born lineared forward modeling data and the observed ones \cite{dutta2014attenuation, zhang2015stable}. Since LSRTM generates simulated data through a linear forward modeling operator with respect to the reflectivity, the simulated data only include primary reflections while neglecting other wavefield terms such as direct arrivals and multiples. In contrast, the observed data contain these higher-order wavefield components, which negatively impacts the misfit measure. In addition, the observed data often correspond to an elastic Earth with attenuating features, and include noise. Thus, designing and selecting an appropriate loss criterion to accurately measure the difference between observed and simulated data is crucial for high-precision reflectivity model imaging \cite{dutta2014cross, yong2019least}. In the past, geophysicists have employed several advanced loss metrics to measure the difference between observed and simulated data, resulting in improved LSRTM imaging. For example, \cite{dutta2014cross} employed a normalized cross-correlation objective function following in the footsteps of the global correlation objective function used in FWI \cite{choi2012application}, which enhances the data matching process by focusing more on the similarity between observed and simulated data while reducing the influence of amplitude. \cite{gu2017elastic} applied a hybrid L1/L2 objective function to improve the robustness of LSRTM when handling noisy data. Later, \cite{yong2019least} utilized the Wasserstein metric as the objective function, which not only accelerates the convergence of LSRTM but also reduces its sensitivity to noise.

In recent years, the rise of machine learning has provided fresh perspectives and opportunities for seismic data analysis and evaluation. Both supervised and unsupervised machine learning techniques have been utilized to develop misfit functions for full-waveform inversion (FWI), achieving favorable inversion results. \cite{sun2022ml} employed meta-learning to train a neural network, resulting in an advanced machine-learning-based loss function that can learn to avoid local minima. However, training such a loss function requires a large and reliable dataset to ensure its applicability to different velocity models. \cite{saad2024siamesefwi} introduced an unsupervised deep learning approach called SiameseFWI to enhance misfit evaluation in FWI \cite{saad2025f}. This method employs a Siamese neural network \cite{melekhov2016siamese} to map observed and simulated data into a transformed representation, where Euclidean distance is applied to compare the extracted features. Compared to traditional FWI, SiameseFWI introduces only minimal computational overhead. In the field of LSRTM, several machine learning-based methods have been proposed for noise removal \cite{wu2024least} and enhancing imaging resolution by estimating the approximate inverse Hessian \cite{kaur2020improving, huang2022fast, sun2023lsmgans}. However, research on applying machine learning to construct the objective function aimed at improving the measurement of observed and simulated data has not yet been explored. 

In this study, we incorporate the Siamese neural network into the LSRTM framework, and we refer to it as SiameseLSRTM. This approach facilitates the evaluation of the misfit between the observed and linear forward modeling data, ultimately enhancing imaging accuracy. The Siamese network comprises two Convolutional Neural Networks (CNNs) with shared weights, with each CNN designed to transform the observed and simulated data into a transformed representation. The network extracts similar features from both the observed and simulated data, allowing the misfit function to effectively measure the differences between them. We apply the proposed SiameseLSRTM to two synthetic datasets and one real land dataset. The imaging results demonstrate that, compared to conventional LSRTM, SiameseLSRTM further improves both imaging accuracy and resolution. In addition, various loss functions are evaluated, including Euclidean loss, L1 loss, and L2 loss, within the SiameseLSRTM framework. The experimental findings indicate that SiameseLSRTM remains effective regardless of the misfit function used. 

The structure of this paper is as follows: We begin by reviewing the fundamental principles of LSRTM, followed by introducing the SiameseLSRTM implementation framework. Next, we present the tests conducted on two synthetic datasets and one real dataset. Finally, we interpret how the Siamese network works to enhance the imaging accuracy of LSRTM and discuss the performance of SiameseLSRTM for noisy data.

\section{Theory}

In this section, we first review the fundamental principles of acoustic LSRTM. Then, we introduce the implementation framework of SiameseLSRTM and explain its working mechanism. Finally, we present the CNN architecture used in the implementation of SiameseLSRTM. 

\subsection{Review of acoustic LSRTM}
LSRTM aims to construct the Earth's reflectivity image by iteratively minimizing the difference between the observed and simulated data \cite{tarantola1984linearized, wong2011least, li2018plane, huang2024least}. LSRTM is often formulated as an optimization problem with an L2 objective function given by:  
\begin{equation}
\label{eq1}
J = \frac{1}{2}\left\| {{d_{obs}} - {d_{sim}}} \right\|_2^2,
\end{equation}
where \({{d_{obs}}}\) and \({{d_{sim}}}\) are the observed and simulated data, respectively. 

Seismic wave forward modeling serves as the foundation for imaging. Here, we use the constant-density acoustic wave equation to simulate seismic wave propagation, which can be expressed as: 
\begin{equation}
\label{eq2}
\frac{1}{{v_0^2(r)}}\frac{{{\partial ^2}{p_0}(r,t)}}{{\partial {t^2}}} - {\nabla ^2}{p_0}(r,t) = f({r_s},t),
\end{equation}
where \({v_0}(r)\) is the background velocity model, \({{p_0}(r,t)}\) is the background wavefield, \({\nabla ^2}\) is the spatial Laplace operator, \(r\) is the spatial coordinate vector, \(t\) is time, and \(f({r_s},t)\) denotes a point source function at the location \({r_s}\). 

For LSRTM, we obtain the simulated data by calculating the perturbation wavefield \(\delta p(r,t)\) associated with the perturbation velocity \(\delta v(r)\). Based on the Born approximation, the equation governing the perturbation wavefield can be written as \cite{Stolt1986migration, plessix2006review, dutta2014attenuation}: 
\begin{equation}
\label{eq3}
\frac{1}{{v_0^2(r)}}\frac{{{\partial ^2}\delta p(r,t)}}{{\partial {t^2}}} - {\nabla ^2}\delta p(r,t) = \frac{{m(r)}}{{v_0^2(r)}}\frac{{{\partial ^2}{p_0}(r,t)}}{{\partial {t^2}}},
\end{equation}
where \({m(r)}\) represents the reflectivity model. To compute the perturbation wavefield and obtain simulated data, we first need to solve equation \ref{eq2} to calculate the background wavefield. Then, we use equation \ref{eq3} to calculate the perturbation wavefield. Unlike equation \ref{eq2}, which employs a point source, equation \ref{eq3} considers the combination of the reflectivity model, background velocity, and background wavefield as a secondary source. 

The gradient calculation formula is often derived using the adjoint state method \cite{lailly1984migration, plessix2006review}, which can be expressed as: 
\begin{equation}
\label{eq4}
g = \frac{{\partial J}}{{\partial m}} = \int_t {\frac{{{\partial ^2}{p_0}(r,t)}}{{\partial {t^2}}}{p_a}} (r,t)dt,
\end{equation}
where \({p_a}(r,t)\) is the adjoint wavefield and can be calculated by
\begin{equation}
\label{eq5}
\frac{1}{{v_0^2(r)}}\frac{{{\partial ^2}{p_a}(r,t)}}{{\partial {t^2}}} - {\nabla ^2}{p_a}(r,t) = \Delta d({r_g},t;{r_s}),
\end{equation}
where \(\Delta d({r_g},t;{r_s})\) denotes the difference between the observed and simulated data at the receiver location \({r_g}\) and for a source location \({r_s}\).

In this study, we utilize the Deepwave package to perform LSRTM \cite{richardson_alan_2023}. In Deepwave, LSRTM is formulated as a recurrent neural network (RNN) and implemented using PyTorch. Unlike traditional gradient computation methods, Deepwave uses automatic differentiation to calculate gradients, which are theoretically equivalent to those obtained using the adjoint state method \cite{richardson2018seismic, sun2020theory}. This RNN-based implementation of LSRTM simplifies programming (requiring only the programming of a forward modeling operator) and facilitates the integration of neural networks into the LSRTM process.

\subsection{The framework of SiameseLSRTM}
\begin{figure*}[!t]
\centering
\includegraphics[width=1\textwidth]{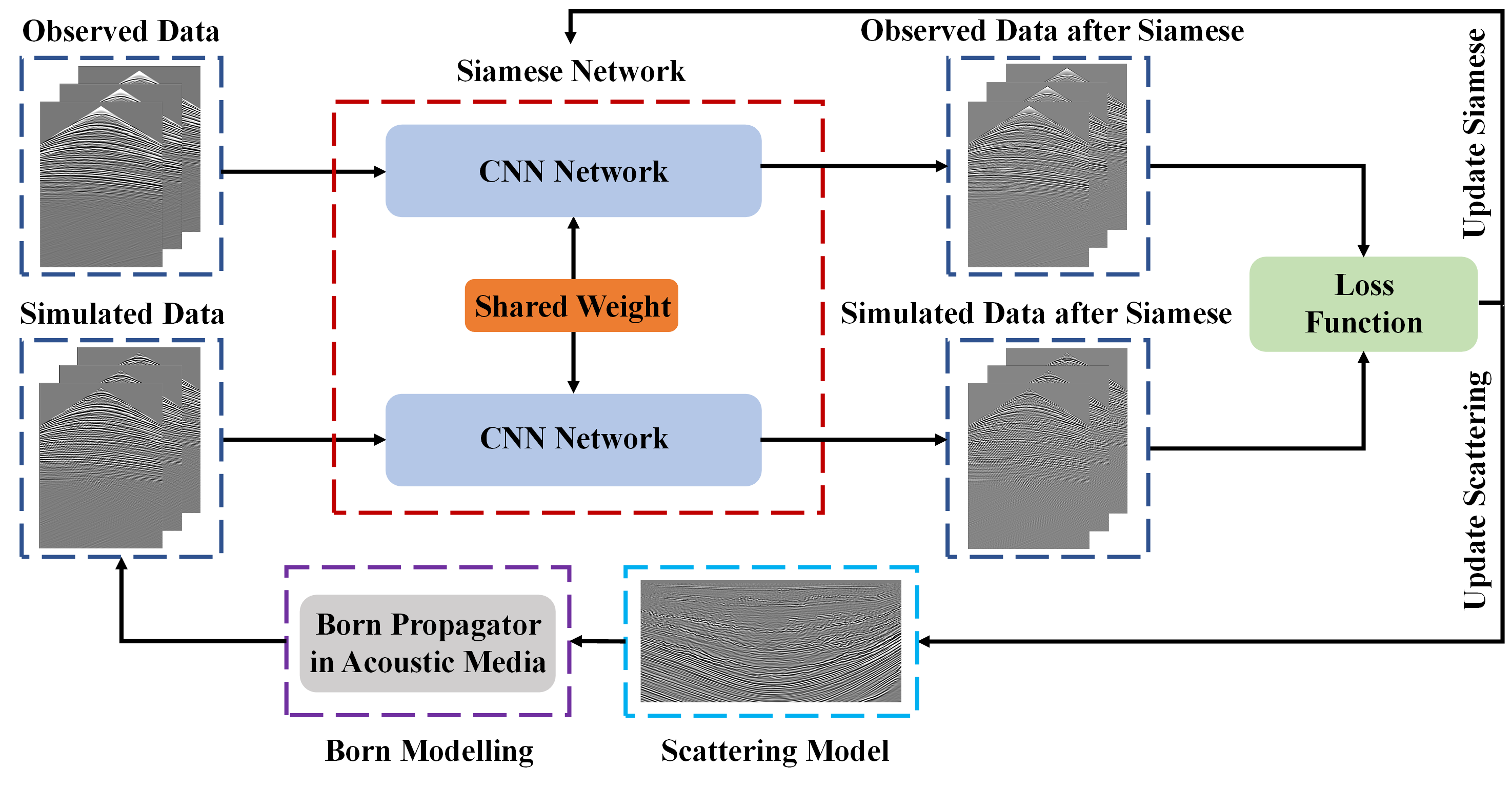}
\caption{A diagram depicting the framework of the proposed SiameseLSRTM. It utilizes a Siamese network comprising two identical CNNs with shared weights, where one CNN processes the observed data, while the other handles the simulated data.}
\label{fig1}
\end{figure*}

LSRTM works by iteratively updating the reflectivity model to minimize the residuals between the observed and simulated data. Therefore, effectively measuring the difference between the observed and simulated data is crucial for improving imaging quality. To enhance this misfit evaluation, we propose the SiameseLSRTM imaging method, with its implementation shown in Figure \ref{fig1}. Compared to traditional LSRTM, SiameseLSRTM employs a Siamese network to map the observed and simulated data into a transformed representation for better comparison. The loss function is then used to measure the difference between the extracted features from the observed and simulated data. The Siamese network consists of two identical CNNs with shared weight parameters. This architecture enables the network to apply the same filters to both the observed and simulated data, allowing for a more precise measurement of the differences between them and enhancing imaging accuracy. The adjoint source is estimated based on the transformed representation through the CNNs, which is then used to update both the reflectivity model and the Siamese network using two Adam optimizers. This ensures that SiameseLSRTM is a self-supervised learning approach.

\subsection{CNN Architecture}
\begin{figure*}[!t]
\centering
\includegraphics[width=1\textwidth]{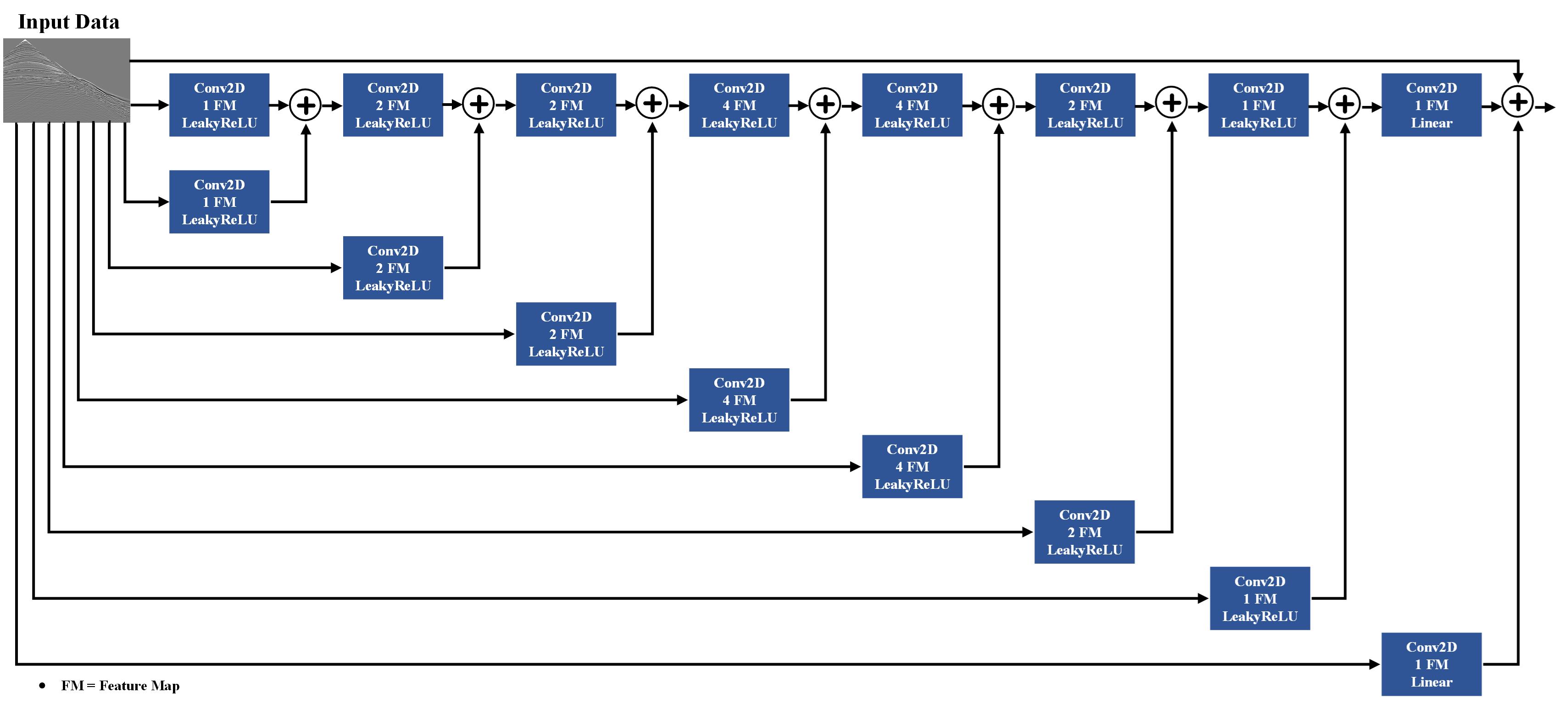}
\caption{The architecture of the CNN used in this study. It consists of eight convolutional layers, each incorporating a skip connection to enhance the network's learning capability.}
\label{fig2}
\end{figure*}

The CNN architecture used here is similar to that in \cite{saad2024siamesefwi}, as shown in Figure \ref{fig2}. The CNN consists of eight 2D convolutional layers, each with a kernel size of 3. Except for the last layer, which uses a linear activation function, all other layers employ the LeakyReLU activation function with a rate of 0.1. The number of feature maps in the eight layers of the CNN is 1, 2, 2, 4, 4, 2, 1, and 1, progressing from shallow to deep. To enhance the learning ability of the CNN, we introduce eight skip connections. Each skip connection starts from the network's input, applies a convolution operation, and directly adds the resulting output to the corresponding convolutional layer. Moreover, the CNN includes a skip connection that provides a direct link from input to output. Thus, the network is really learning the required changes to the data to improve the comparison. This skip connection, also, helps prevent the output from becoming random noise due to the network's parameter initialization in the early iterations. Consequently, the proposed framework operates as a regular LSRTM for the first few epochs until it begins to optimize its parameters to learn to improve the data matching, facilitating smooth and stable LSRTM. 

\section{NUMERICAL EXPERIMENTS}
In this section, we validate the effectiveness of the proposed SiameseLSRTM in improving imaging accuracy through tests on the SEAM and Marmousi models, as well as a field dataset from a land seismic survey. These tests demonstrate that, compared to LSRTM, SiameseLSRTM can further suppress imaging noise, improve resolution, and balance amplitude. Note that all experiments were conducted on a single NVIDIA A100 GPU with 80 GB of memory.

\subsection{SEAM model}
\begin{figure*}
\centering
\includegraphics[width=0.7\textwidth]{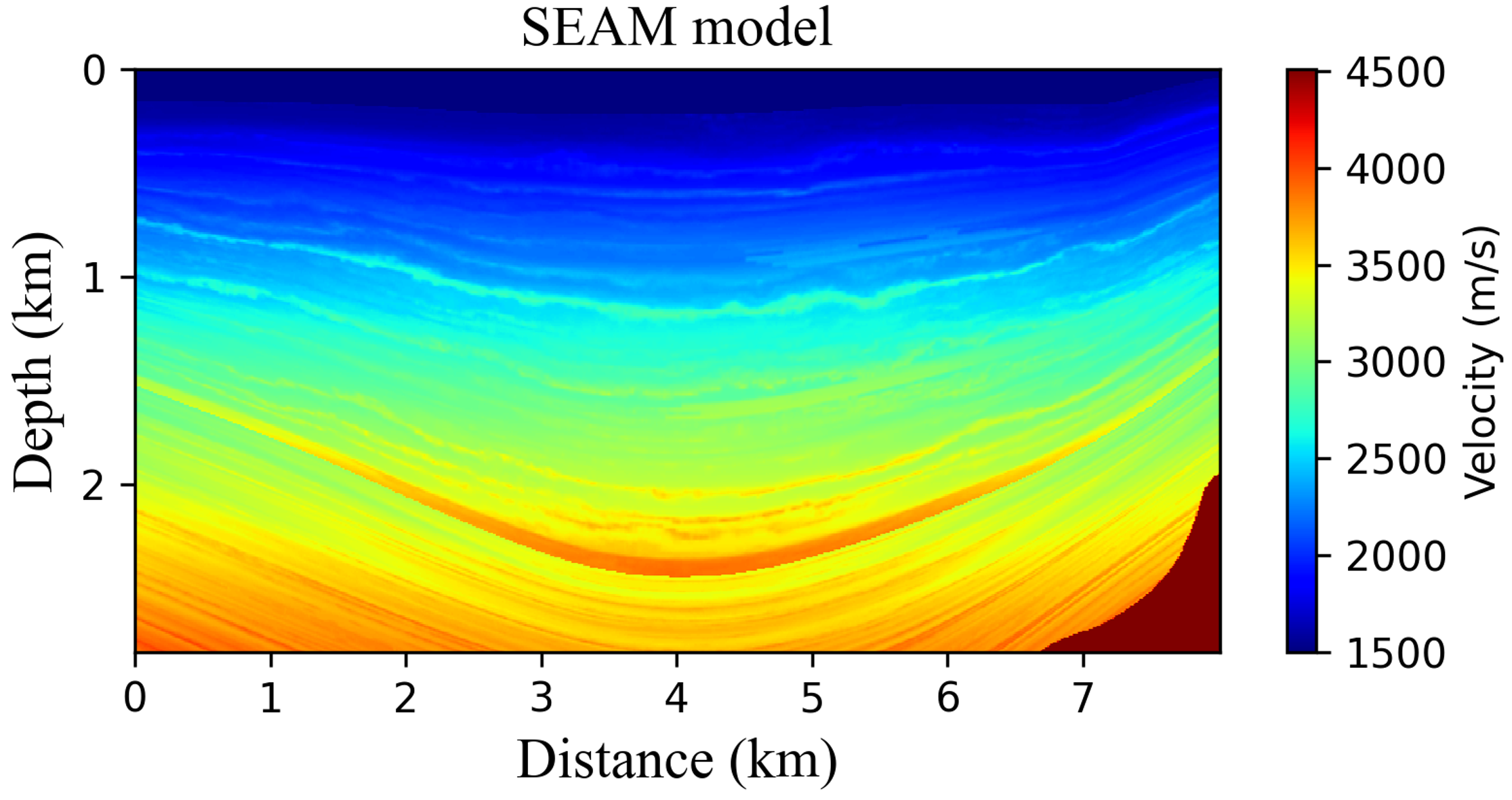}
\caption{P-wave velocity of the SEAM model.}
\label{fig3}
\end{figure*}

Figure \ref{fig3} shows a part of the SEAM Phase I RPSEA Model \cite{fehler2008seg}, which features thin layers. This model is utilized to validate the effectiveness of the developed SiameseLSRTM in imaging thin layers. The model contains a grid size of 1001 × 351, with grid spacing of 8 m in both the horizontal and vertical directions. A 25 Hz Ricker wavelet is used as the source. The first shot is positioned 1400 m from the left end of the model, followed by 65 additional sources evenly spaced at 80 m intervals along the surface from left to right. A total of 351 receivers are used, evenly distributed 
with a spacing of 8 m, and all receivers are located on the surface. The time sampling interval is 2 ms, and the total recording duration is 4 s. Equation \ref{eq2}, with the true velocity, is used to generate the observed records, employing a numerical discretization that achieves second-order accuracy in time and eighth-order accuracy in space. We mute the direct arrivals before performing LSRTM imaging. To demonstrate the effectiveness of SiameseLSRTM regardless of the loss function, we conduct tests based on Euclidean loss, L2 loss, and L1 loss for both LSRTM and SiameseLSRTM. For LSRTM, the learning rates for the image optimizer are set to 6 for both Euclidean loss and L2 loss, and 10 for L1 loss. For SiameseLSRTM, the learning rates for the image optimizer are the same as those used in LSRTM. For Euclidean loss, L2 loss, and L1 loss, the learning rates for the Siamese network optimizer are set to 1e-3, 5e-4, and 1e-3, respectively.

\begin{figure*}
\centering
\includegraphics[width=1\textwidth]{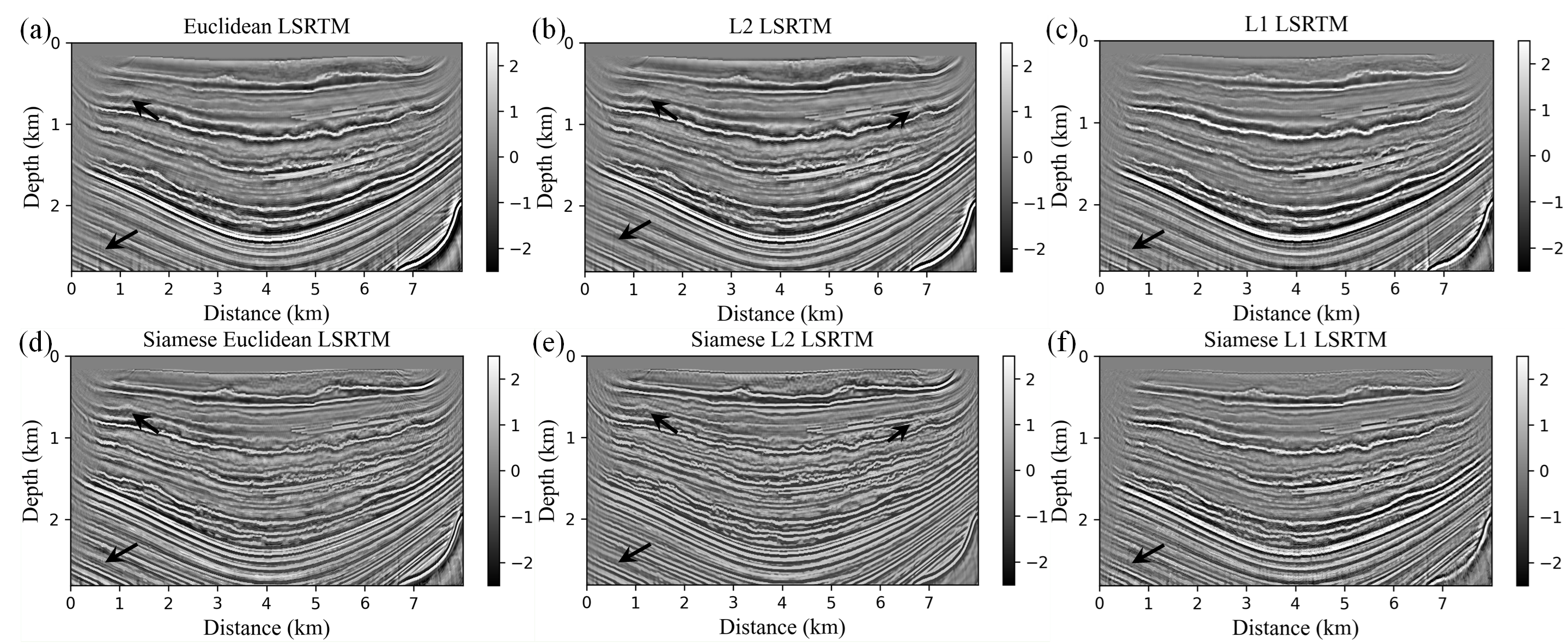}
\caption{Migration results after 20 iterations. The first row displays the LSRTM imaging results based on (a) Euclidean loss, (b) L2 loss, and (c) L1 loss. The second row shows the SiameseLSRTM imaging results, also based on (d) Euclidean loss, (e) L2 loss, and (f) L1 loss.}
\label{fig4}
\end{figure*}

\begin{figure*}
\centering
\includegraphics[width=1\textwidth]{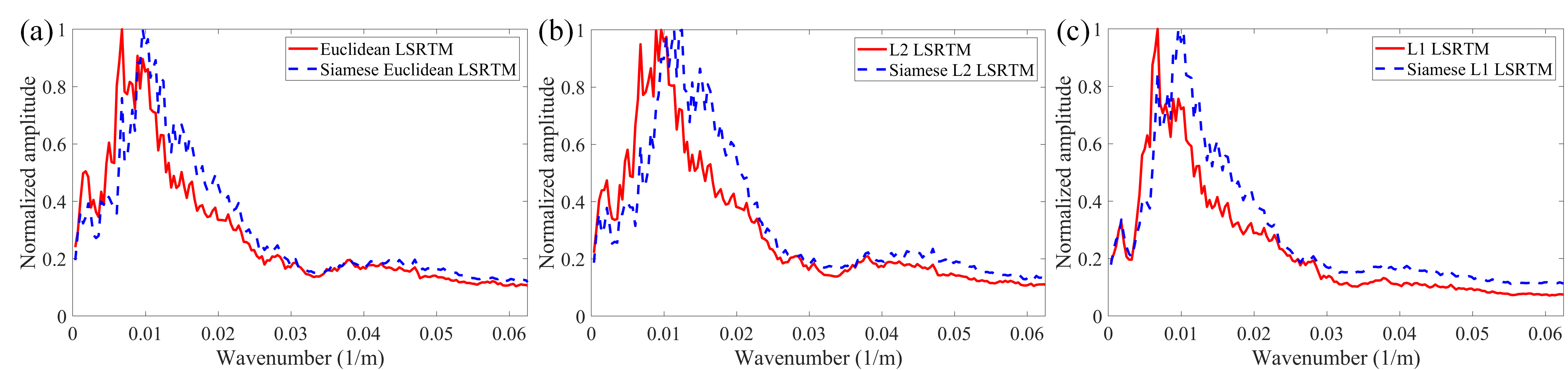}
\caption{Frequency spectra corresponding to the imaging results in Figure \ref{fig4}. The red and blue lines represent the frequency spectra of the LSRTM and Siamese LSRTM imaging results, respectively. Panels (a), (b), and (c) display the frequency spectra obtained from imaging results based on Euclidean loss, L2 loss, and L1 loss, respectively.}
\label{fig5}
\end{figure*}

Figure \ref{fig4} shows the imaging results of LSRTM and SiameseLSRTM after 20 iterations, based on different loss functions. The first and second rows of Figure \ref{fig4} show the imaging results for LSRTM and SiameseLSRTM, respectively. The first to third columns of Figure \ref{fig4} show the results obtained using Euclidean loss, L2 loss, and L1 loss. As indicated by the arrows in Figure \ref{fig4}, SiameseLSRTM further suppresses imaging noise compared to LSRTM and is effective for different loss functions.  We also plot the frequency spectra of all imaging results in Figure \ref{fig5}. The red lines represent the frequency spectra calculated from the LSRTM imaging results, while the blue lines represent the frequency spectra calculated from the SiameseLSRTM imaging results. From this, we observe that the blue lines shift toward higher frequencies compared to the red lines, indicating that SiameseLSRTM further enhances imaging resolution over LSRTM.

\begin{figure*}
\centering
\includegraphics[width=1\textwidth]{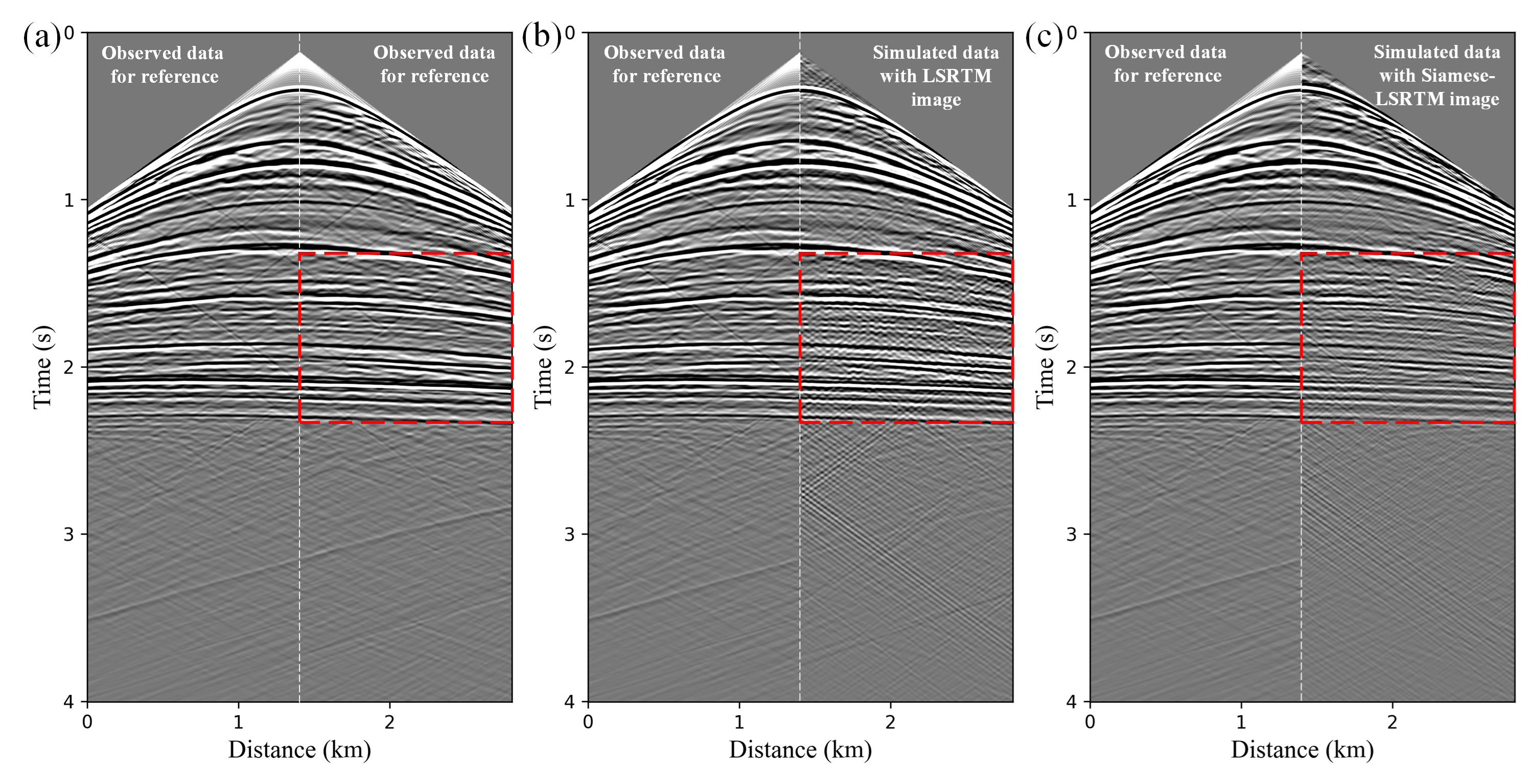}
\caption{Comparison of observed and simulated data for the 30th shot. Panel (a) shows the observed data for comparison purposes. The simulated data in panels (b) and (c) is calculated based on Figures \ref{fig4}(a) and \ref{fig4}(d), respectively. The simulated shot gathers are computed using equation \ref{eq3}. The amplitudes in all four panels are clipped to the same level.}
\label{fig6}
\end{figure*}

We also compare the observed and simulated data in Figure \ref{fig6}. In the regions highlighted by the red rectangle boxes in Figure \ref{fig6}, the simulated data generated using the LSRTM imaging result (Figure \ref{fig6}b) exhibits noise when compared to the observed data (Figure \ref{fig6}a), indicating the presence of migration noise in the LSRTM result. In contrast, the data simulated using the SiameseLSRTM imaging result (Figure \ref{fig6}c) is free from noise and closely resembles the observed data (Figure \ref{fig6}a). This demonstrates that SiameseLSRTM effectively suppresses imaging noise more effectively than LSRTM.

Table \ref{tab1} shares the running time required to image the SEAM model using different algorithms. Compared to LSRTM, SiameseLSRTM results in only a slight increase in computational cost, increasing by approximately 1\%.

\begin{table}[htbp]
\centering
\caption{Comparison of the elapsed time for the SEAM and Marmousi model tests using different algorithms.}
\footnotesize
\begin{tabular}{ccccccc}
\hline
Algorithm       & \begin{tabular}[c]{@{}c@{}}Euclidean \\ LSRTM\end{tabular} & \begin{tabular}[c]{@{}c@{}}Siamese \\ Euclidean LSRTM\end{tabular} & \begin{tabular}[c]{@{}c@{}}L2 \\ LSRTM\end{tabular} & \begin{tabular}[c]{@{}c@{}}Siamese \\ L2 LSRTM\end{tabular} & \begin{tabular}[c]{@{}c@{}}L1 \\ LSRTM\end{tabular} & \begin{tabular}[c]{@{}c@{}}Siamese \\ L1 LSRTM\end{tabular} \\ \hline
\begin{tabular}[c]{@{}c@{}}Running time of \\ SEAM model (s)\end{tabular}    & 470                                                        & 490                                                                & 470                                                 & 490                                                         & 468                                                 & 490                                                         \\ \hline
\begin{tabular}[c]{@{}c@{}}Running time of \\ Marmousi model (s)\end{tabular} & 697                                                        & 729                                                                & 695                                                 & 729                                                         & 697                                                 & 731                                                         \\ \hline
\end{tabular}
\label{tab1}
\end{table}

\subsection{Marmousi model}
\begin{figure*}
\centering
\includegraphics[width=0.7\textwidth]{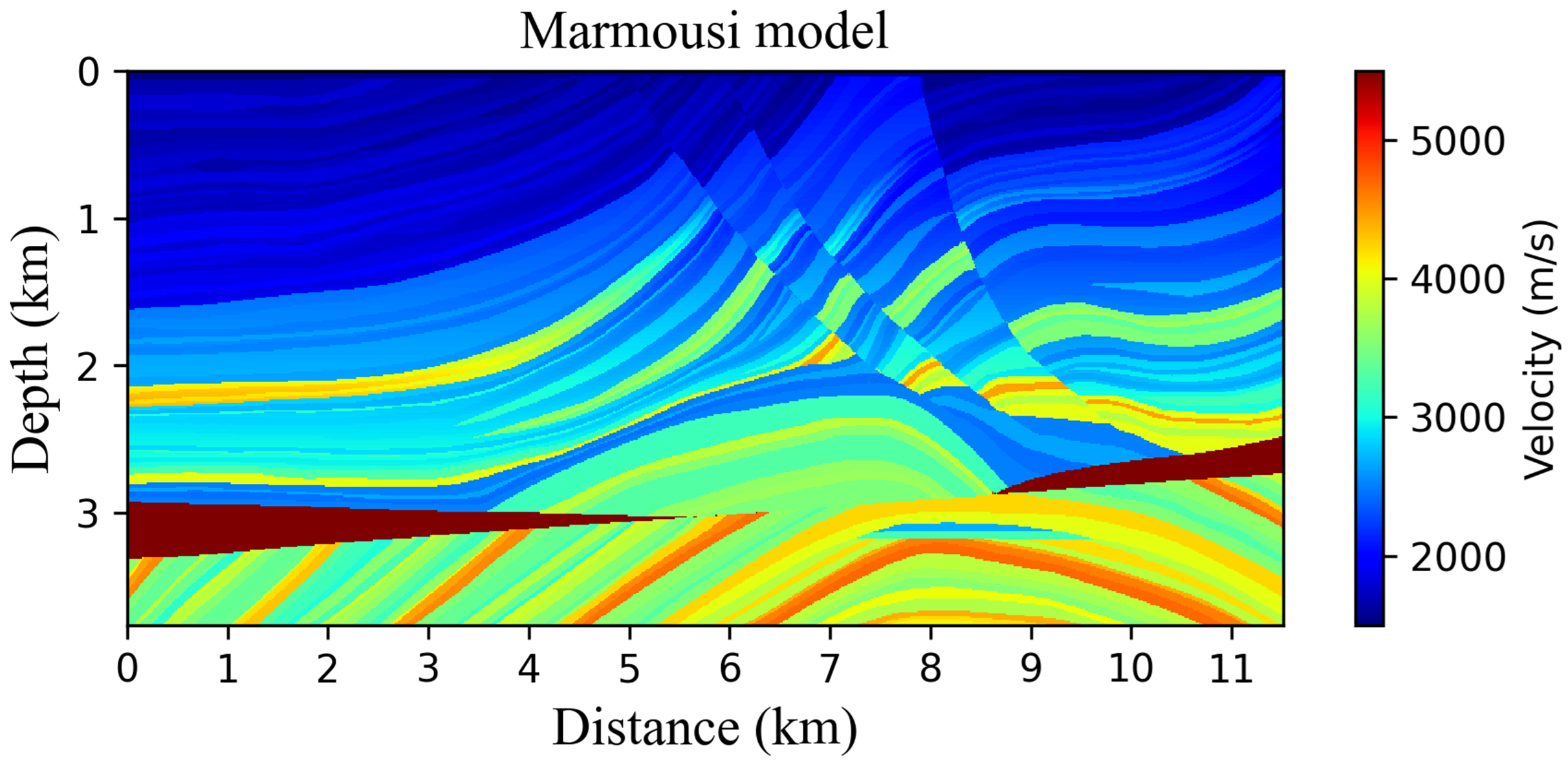}
\caption{P-wave velocity of the Marmousi model.}
\label{fig7}
\end{figure*}

We conduct a validation test of SiameseLSRTM using the complex Marmousi model, as shown in Figure \ref{fig7} \cite{versteeg1991practical}, which features faults and steeply dipping layers. The Marmousi model has a grid size of 1151 × 376, with a uniform grid spacing of 10 m in both the vertical and horizontal directions. A 20 Hz Ricker wavelet with a 2 ms time sampling interval is used as the source to generate 4.4 s of data. A total of 77 shots are excitated, with the first shot located 1880 m from the left end of the model. The remaining shots are evenly spaced at 100 m intervals from left to right. Each shot is recorded by 176 receivers, symmetrically distributed on both sides of the source, with a receiver spacing of 10 m. Both the sources and receivers are positioned at the surface. Similarly, direct wave removal is performed before applying LSRTM and SiameseLSRTM. Here, we also employ Euclidean loss, L2 loss, and L1 loss to demonstrate that SiameseLSRTM is both effective and robust across different loss functions. In this case, for both LSRTM and SiameseLSRTM with different loss functions, the learning rate of the image optimizer is set to 30. For SiameseLSRTM, the learning rates of the Siamese network corresponding to Euclidean loss, L2 loss, and L1 loss are set to 2e-3, 8e-4, and 1e-3, respectively.

\begin{figure*}
\centering
\includegraphics[width=1\textwidth]{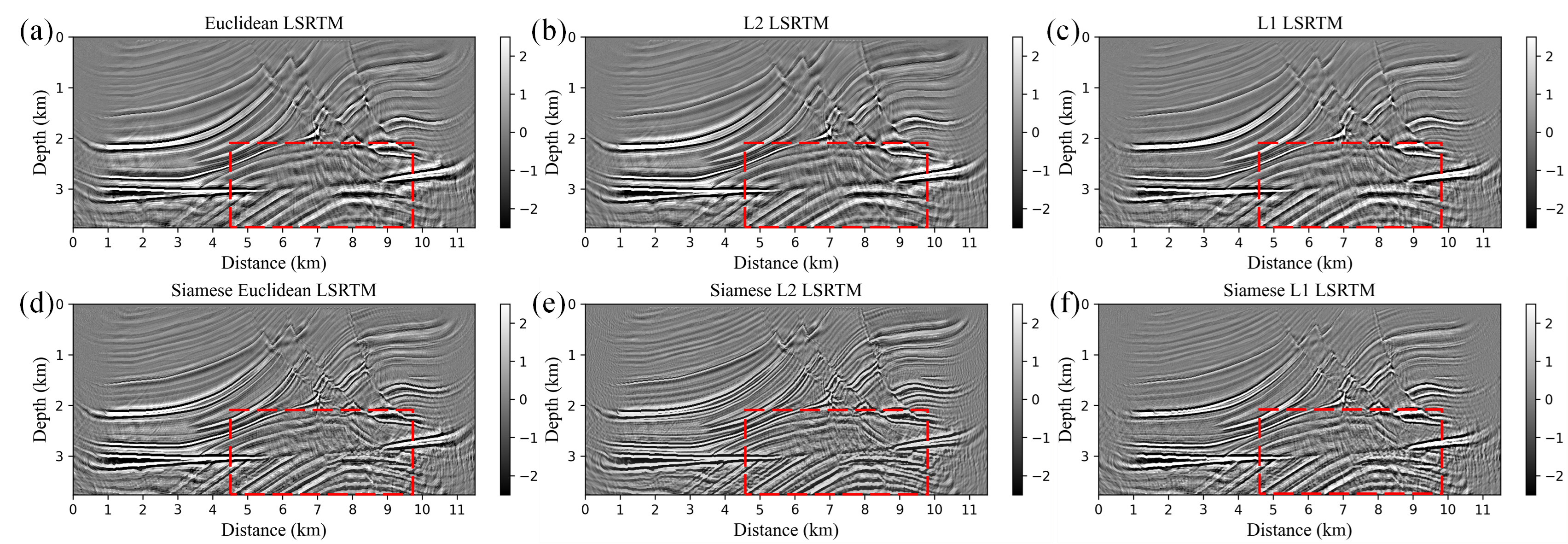}
\caption{Migration imaging results after 20 iterations. (a)–(c) show LSRTM imaging results using Euclidean loss, L2 loss, and L1 loss, respectively. (d)–(f) show SiameseLSRTM imaging results using Euclidean loss, L2 loss, and L1 loss, respectively.}
\label{fig8}
\end{figure*}

\begin{figure*}
\centering
\includegraphics[width=1\textwidth]{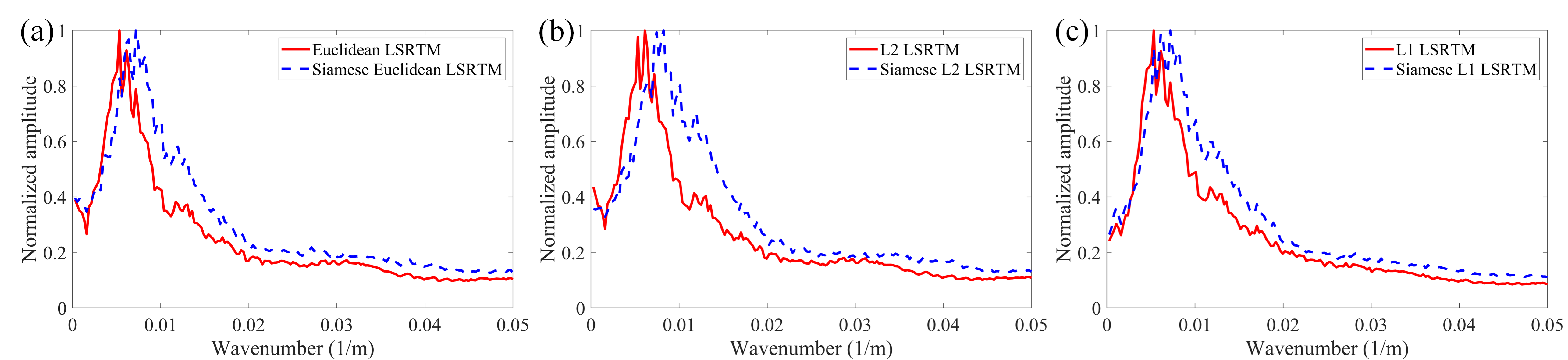}
\caption{Frequency spectra computed from the imaging results in Figure \ref{fig8}. The red and blue lines represent the frequency spectra of LSRTM and SiameseLSRTM imaging results, respectively. Panels (a), (b), and (c) show the spectra corresponding to imaging results obtained using Euclidean loss, L2 loss, and L1 loss, respectively.}
\label{fig9}
\end{figure*}

Figure \ref{fig8} shows the imaging results of LSRTM and SiameseLSRTM using different loss functions. By comparing the LSRTM and Siamese LSRTM imaging results, it is clear that SiameseLSRTM can further remove low-wavenumber artifacts caused by backscattering \cite{liu2011effective}, in contrast to LSRTM. As shown in the red rectangle boxes in Figure \ref{fig8}, SiameseLSRTM can further suppress arc-shape noise compared to LSRTM, producing a clearer image of the subsurface structure. Additionally, we find that SiameseLSRTM is effective with different loss functions, all of which improve imaging accuracy. We plot the frequency spectra of the imaging results from Figure \ref{fig8} in Figure \ref{fig9}. The red lines represent the frequency spectra of the LSRTM imaging results, while the blue lines represent the frequency spectra of the SiameseLSRTM imaging results. It is clear that the peaks of the blue lines shift towards higher frequencies compared to the red lines, indicating that SiameseLSRTM can further improve imaging resolution compared to LSRTM.

\begin{figure*}
\centering
\includegraphics[width=1\textwidth]{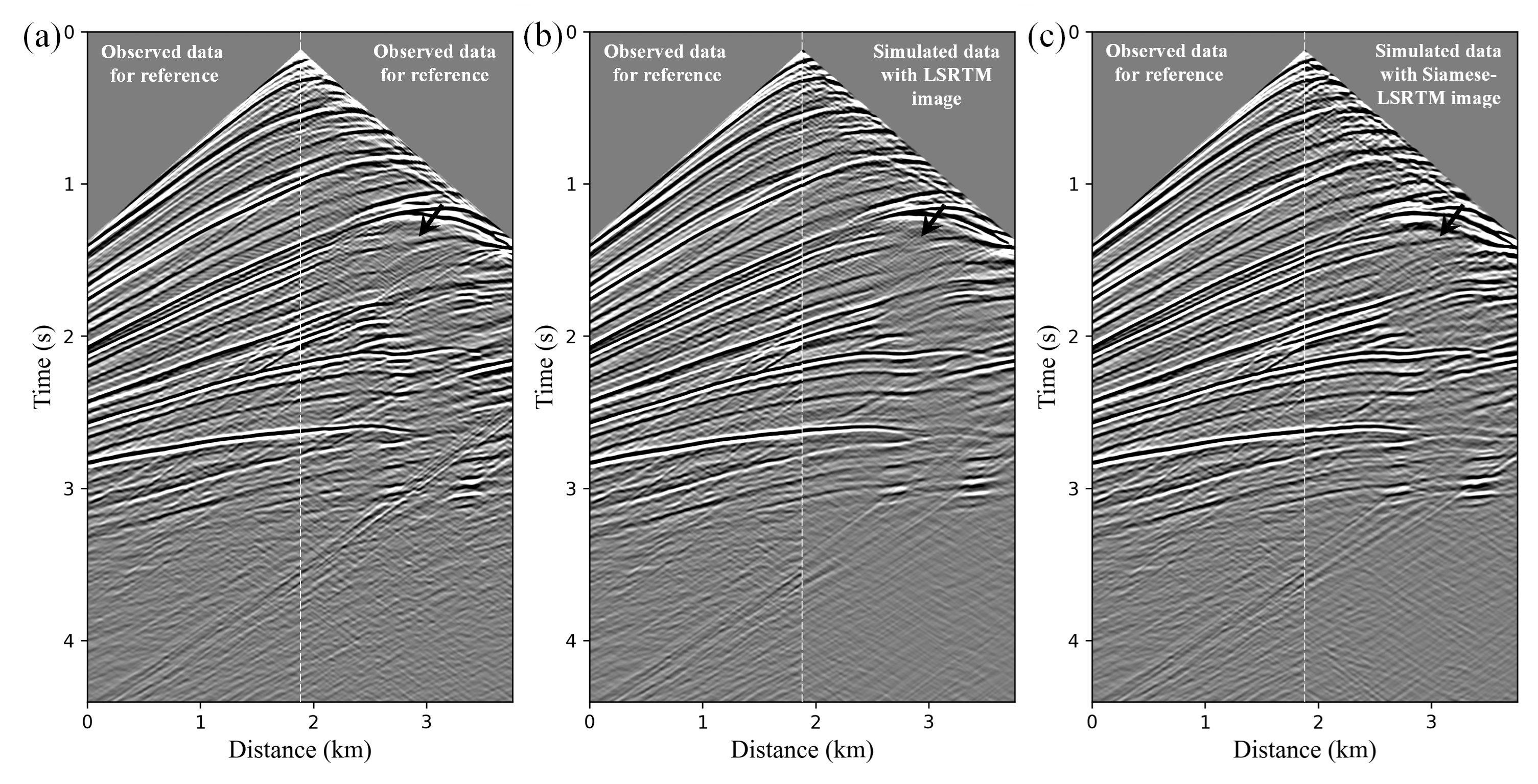}
\caption{Comparison between observed data and simulated data generated using LSRTM images after the 20th iteration. (a) shows the observed data, while (b) and (c) display the data simulated using LSRTM (Figure \ref{fig8}a) and SiameseLSRTM (Figure \ref{fig8}d) images, respectively. The shot number is 30, and the amplitudes in all panels are clipped to the same level.}
\label{fig10}
\end{figure*}

We also generate the simulated data using the imaging results from Figure \ref{fig8}a and \ref{fig8}d, and compare them with the observed data (shown in Figure \ref{fig10}). We notice that the shot records simulated using the SiameseLSRTM imaging result are closer to the observed records in terms of energy, particularly for weak signals. This indicates that SiameseLSRTM better balances imaging amplitudes. Additionally, as indicated by the black arrows in Figure \ref{fig10}, the data simulated from the LSRTM imaging result shows an energy gap, leading to discontinuities in the seismic event. In contrast, the data simulated from SiameseLSRTM result yields a continuous seismic event. This further demonstrates that SiameseLSRTM, compared to LSRTM, can obtain more balanced imaging results and produces high-accuracy imaging results.

Table \ref{tab1} shows the running time for imaging the Marmousi model using LSRTM and SiameseLSRTM. Similarly, we observe that SiameseLSRTM introduces only approximately 1\% increase in computational cost compared to LSRTM. This demonstrates that SiameseLSRTM achieves improved imaging accuracy with only a slight increase in computational cost.


\subsection{Land field data}
\begin{figure*}
\centering
\includegraphics[width=0.7\textwidth]{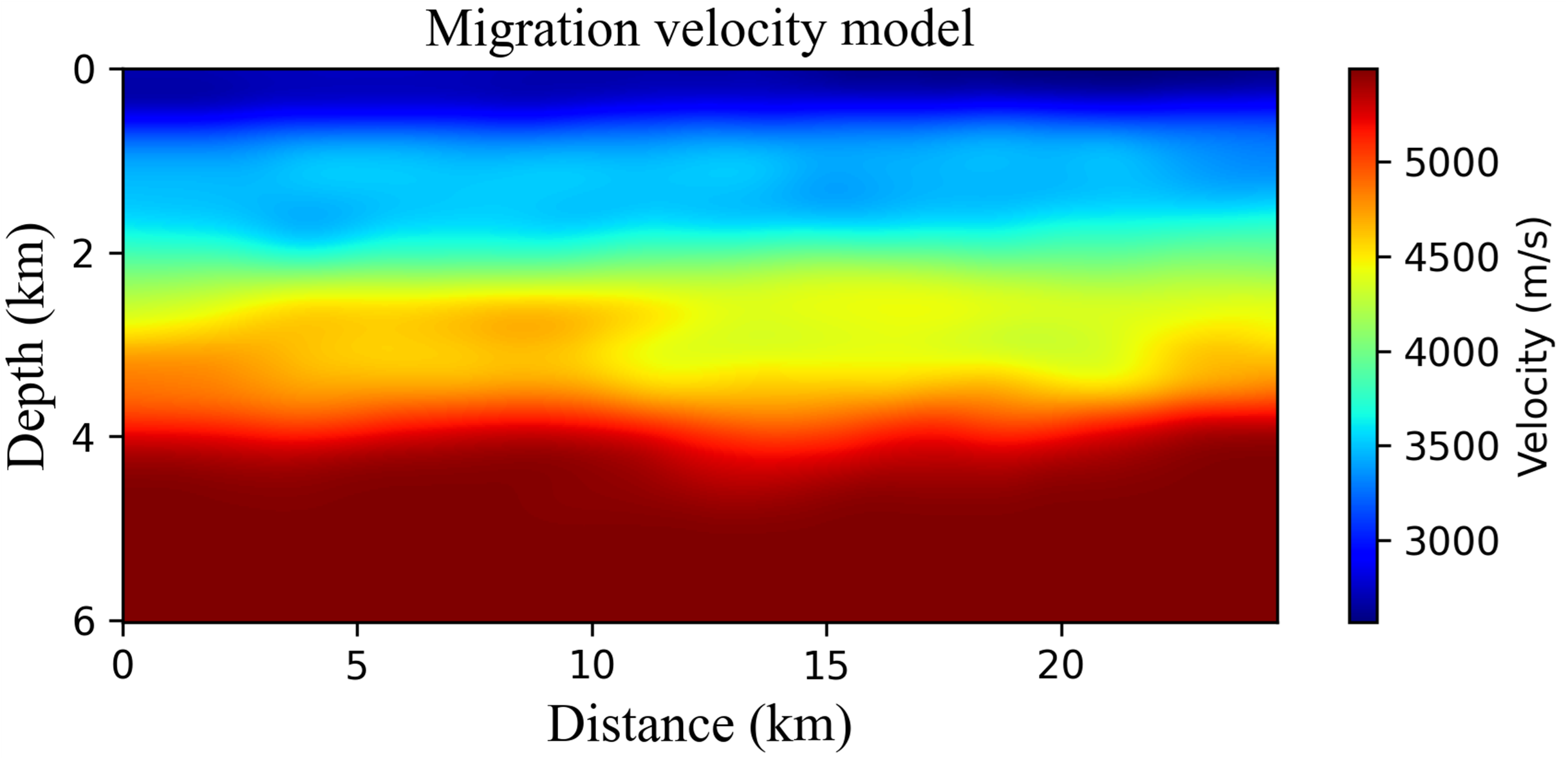}
\caption{Migration velocity model for the land seismic data.}
\label{fig11}
\end{figure*}

In this section, we use a real land dataset to validate the effectiveness of the proposed SiameseLSRTM in improving migration imaging accuracy. The migration velocity model is shown in Figure \ref{fig11}, with a grid size of 1472 × 301 and a grid spacing of 20 m in both the vertical and horizontal directions. The dataset includes 63 shots, evenly distributed with a 350 m shot interval. Each shot is recorded with 140 receivers spaced 50 m apart. The total recording duration is 3.6 s, and the resampled time sampling interval is 0.0006 s. A Ricker wavelet with a peak frequency of 20 Hz is used as the seismic source. 

\begin{figure*}
\centering
\includegraphics[width=\textwidth]{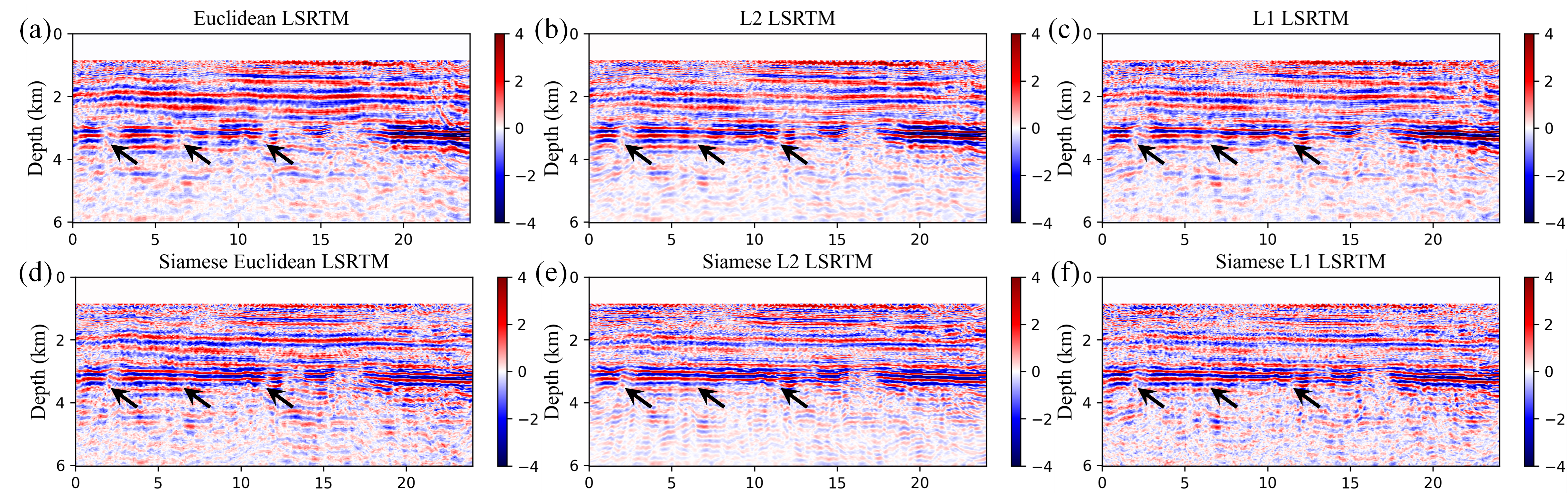}
\caption{Imaging results of LSRTM using (a) Euclidean loss, (b) L2 loss, and (c) L1 loss; Imaging results of SiameseLSRTM using (d) Euclidean loss, (e) L2 loss, and (f) L1 loss. Clear improvements are highlighted by the black arrows.}
\label{fig12}
\end{figure*}

\begin{figure*}
\centering
\includegraphics[width=1\textwidth]{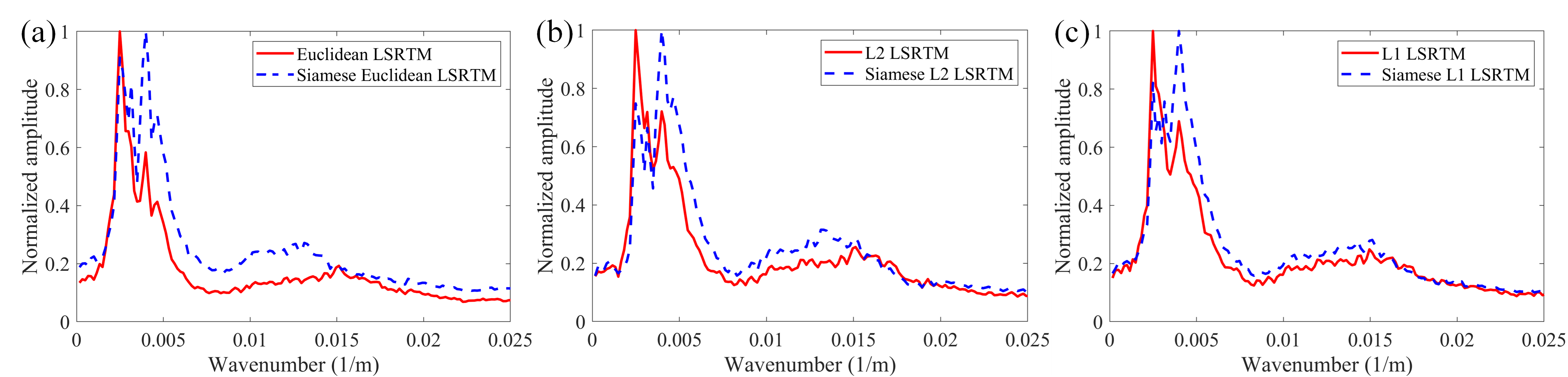}
\caption{Frequency spectra for the LSRTM and SiameseLSRTM images shown in Figure \ref{fig12}.}
\label{fig13}
\end{figure*}

Figure \ref{fig12} shows the imaging results of LSRTM and SiameseLSRTM after 20 iterations using different loss functions. Compared to LSRTM, the imaging results from SiameseLSRTM exhibit more continuous seismic events (as indicated by the black arrows), better amplitude balance in the mid-to-deep regions, and enhanced resolution. Furthermore, regardless of the loss function used, SiameseLSRTM consistently achieves higher imaging accuracy than LSRTM. Figure \ref{fig13} shows the frequency spectra of the imaging results corresponding to those in Figure \ref{fig12}, where it is evident that the imaging results from SiameseLSRTM exhibit a broader frequency band compared to those from LSRTM. This demonstrates that the SiameseLSRTM can further enhance the imaging resolution compared to the traditional LSRTM. 

\begin{figure*}
\centering
\includegraphics[width=1\textwidth]{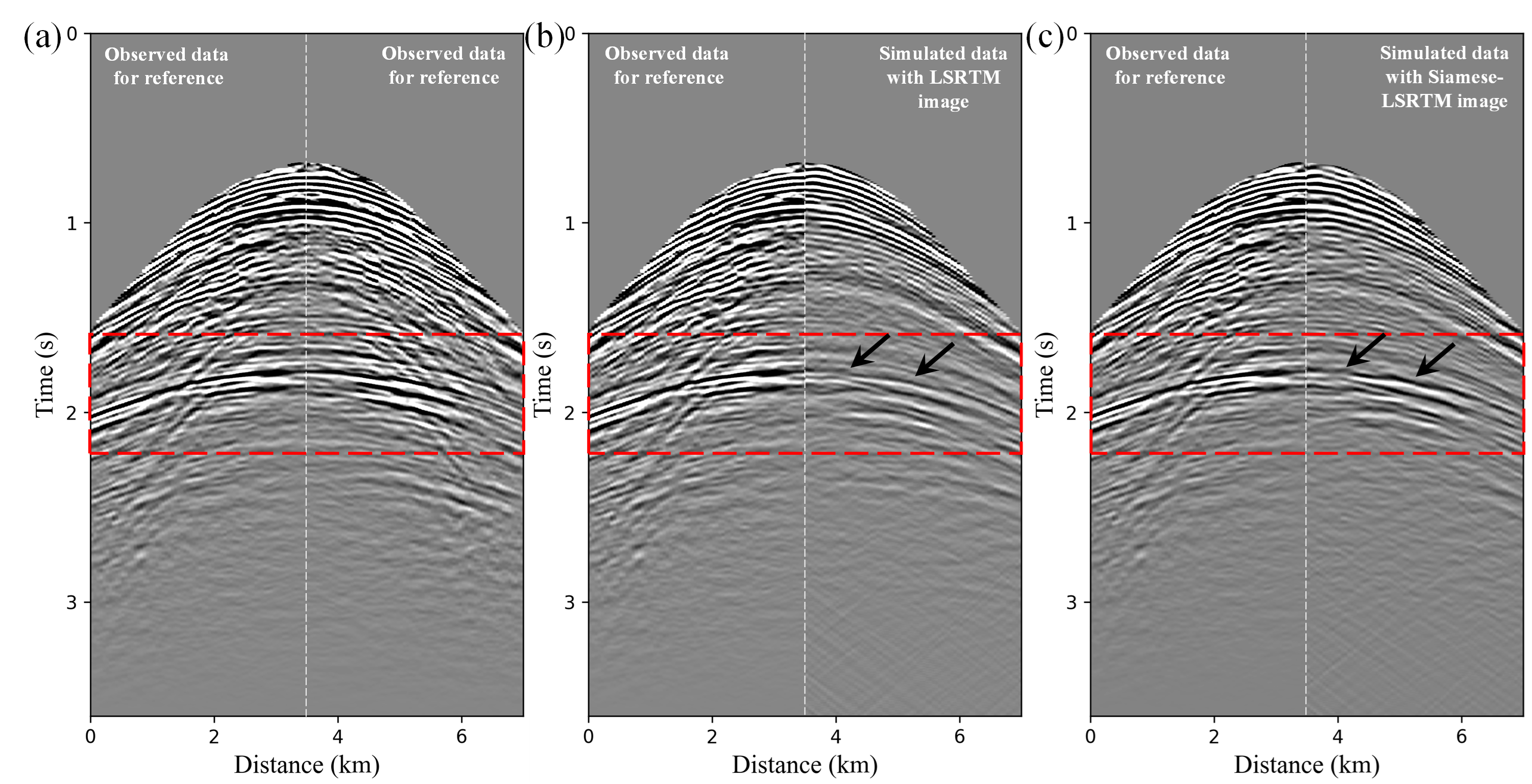}
\caption{Comparison between the observed and the simulated data generated from the imaging results shown in Figures \ref{fig12}(a) and \ref{fig12}(d). Note that all panels are plotted using the same percentile parameters for clipping.}
\label{fig14}
\end{figure*}

Figure \ref{fig14} shows a comparison between the observed and simulated data. The simulated data in Figure \ref{fig14}b and Figure \ref{fig14}c are generated using LSRTM image (Figure \ref{fig12}a) and SiameseLSRTM image (Figure \ref{fig12}d), respectively. From Figure \ref{fig14}, it is clear that the simulated data generated using the SiameseLSRTM image aligns more closely with the observed data compared to those simulated using the LSRTM image (as indicated by the red rectangular boxes and black arrows). This suggests that the imaging results from SiameseLSRTM are more accurate than those from LSRTM. 

\section{DISCUSSION}

In this section, we first explain the essence of using the Siamese network to improve imaging accuracy in LSRTM. Then, we conduct noise sensitivity tests on SiameseLSRTM. Finally, we share strategies for selecting the learning rate.

\subsection{Interpretation of the performance of the Siamese network}
\begin{figure*}
\centering
\includegraphics[width=0.8\textwidth]{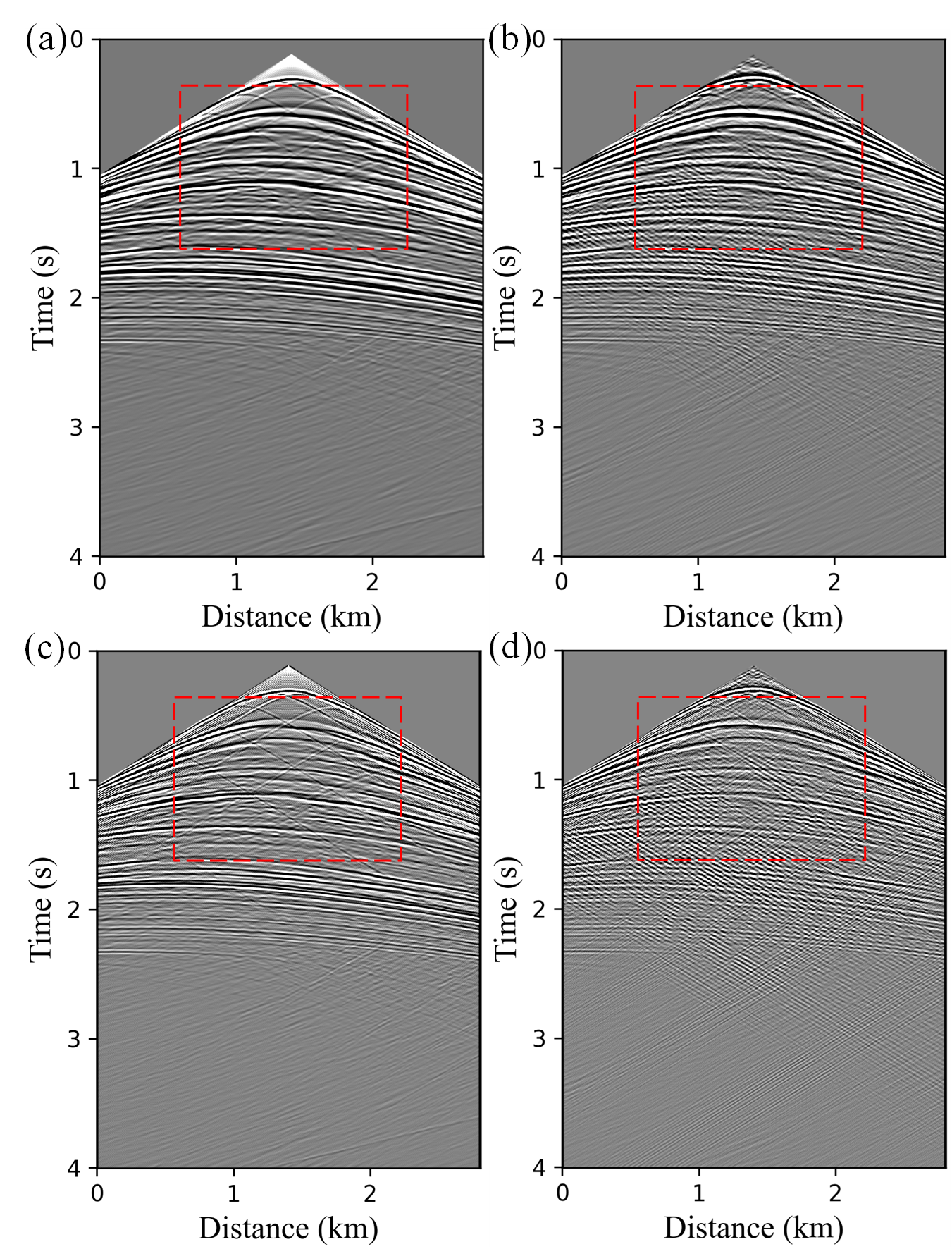}
\caption{Comparison of shot gathers before and after inputting into the Siamese network. (a) Observed data; (b) Simulated data generated from the imaging result in Figure \ref{fig4}d; (c) the output of the Siamese network for input (a); (d) the output of the Siamese network for input (b). Note that all panels are plotted using the same percentile parameters for clipping.}
\label{fig15}
\end{figure*}

\begin{figure*}
\centering
\includegraphics[width=1\textwidth]{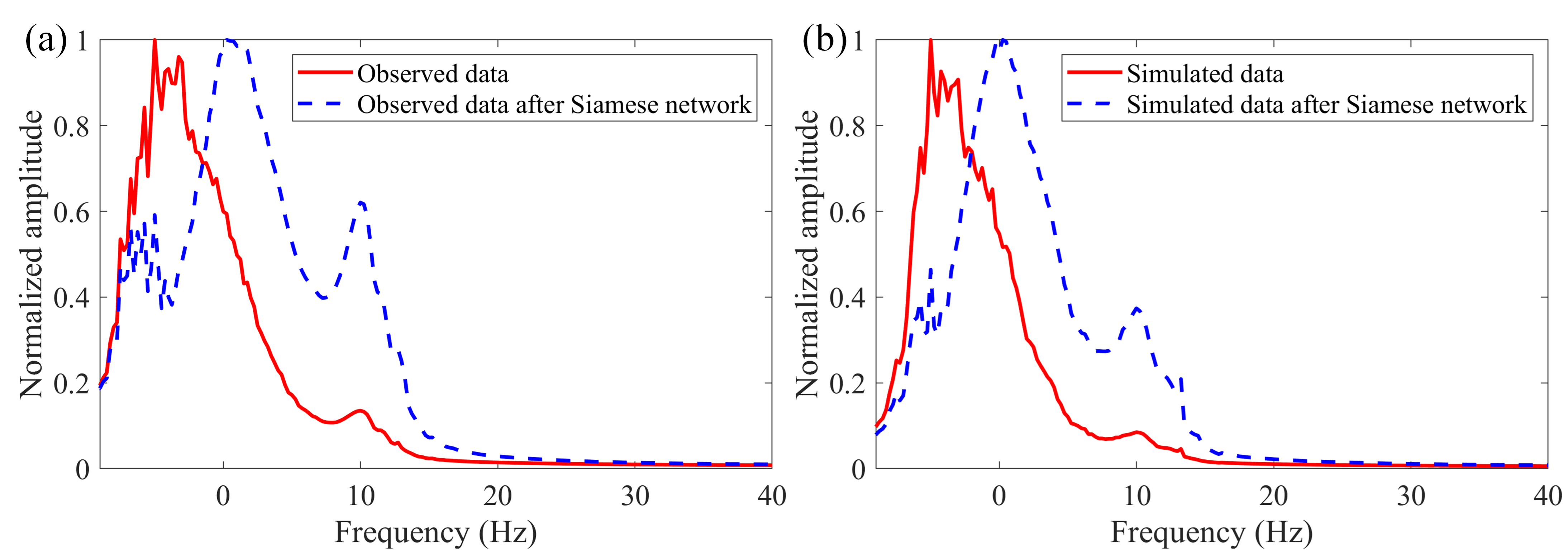}
\caption{Spectra of the region highlighted by the red rectangle boxes in Figure \ref{fig15}. (a) and (b) show the spectra of the observed and simulated data, respectively. The red and blue lines represent the spectra of the data before and after passing through the Siamese network, respectively.}
\label{fig16}
\end{figure*}

Here, we use the SEAM model as an example to interpret the performance of the Siamese network in SiameseLSRTM. Figures \ref{fig15}a and \ref{fig15}c show the observed data before and after being processed by the Siamese network at the 20th iteration in the SEAM model test presented in Figure \ref{fig4}. As highlighted by the red rectangle boxes, the weak reflection signals in the observed data are enhanced after processing by the Siamese network, leading to better amplitude balance between strong and weak signals.  Figures \ref{fig15}b and \ref{fig15}d display the simulated data before and after being processed by the Siamese network at the 20th iteration. Similarly, as indicated by the red rectangle boxes, the weak reflection signals in the simulated data are also enhanced after processing by the Siamese network. Since the weak reflection signals in both the observed and simulated data are enhanced, SiameseLSRTM achieves better amplitude balancing than LSRTM, as highlighted by the synthetic and field data test. With the amplification of weak reflection signals, the reflection energy in the imaging results also increases, while noise is correspondingly suppressed. We also plot the frequency spectra of the region highlighted by the red rectangle boxes in Figure \ref{fig15} and display them in Figure \ref{fig16}. From Figure \ref{fig16}, we observe that the dominant frequency of the data after processing through the Siamese network shifts toward higher frequencies, indicating an improvement in the resolution of the data. This explains why SiameseLSRTM enhances imaging resolution. Since the Siamese network consists of two CNNs with shared weights, its primary function is to extract or emphasize signals with common characteristics from both the observed and simulated data. The observed data contain reflections, multiples, and diffractions, while the simulated data, generated through a linear forward modeling operator, only include reflections. Therefore, the common component between the observed and simulated data is the reflection signal. This explains why we observe an enhancement of reflection signals in both the observed and simulated datasets after processing through the Siamese network. This enhancement plays a crucial role in measuring the differences between the observed and simulated data, ultimately improving the imaging accuracy of LSRTM.

\subsection{The Sensitivity of the SiameseLSRTM to the noise}
\begin{figure*}
\centering
\includegraphics[width=1\textwidth]{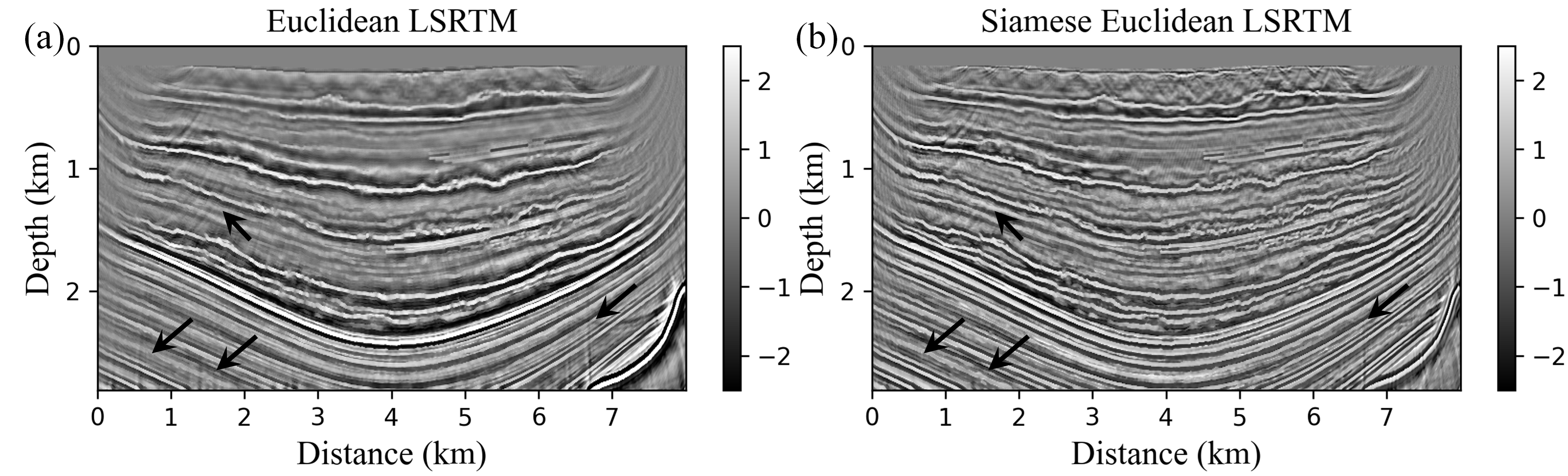}
\caption{LSRTM (a) and SiameseLSRTM (b) imaging results using Euclidean loss for noisy data with SNR = 5.}
\label{fig17}
\end{figure*}

To evaluate the effectiveness of SiameseLSRTM in imaging noisy data, we use the SEAM model as an example. We add white Gaussian noise (AWGN) to the observed data, resulting in a signal-to-noise ratio (SNR) of 5 dB for the noisy data. Figure \ref{fig17} shows the imaging results obtained using LSRTM and SiameseLSRTM for noisy data. The results demonstrate that both LSRTM and SiameseLSRTM maintain high-accuracy in imaging despite the presence of noise. Additionally, as indicated by the black arrows in Figure \ref{fig17}, SiameseLSRTM effectively suppresses arc-shaped imaging noise. This indicates that SiameseLSRTM exhibits strong robustness and stability in imaging noisy data.

\subsection{Strategies for learning rate determination}
The SiameseLSRTM employs two Adam optimizers: one to update the velocity model and another to update the CNN. Consequently, selecting appropriate learning rates is crucial for achieving high-precision imaging results with SiameseLSRTM. Due to skip connections from input to output in the CNN, a near-zero learning rate for the CNN causes SiameseLSRTM to revert to standard LSRTM. Conversely, an excessively high CNN learning rate may result in rapid updates that compromise imaging accuracy. Thus, a small learning rate, typically on the order of 1e-4 or 1e-3 , is preferred, ensuring that the CNN updates in the intended direction while allowing sufficient time for effective refinement of the velocity model.


\section{Conclusion}
We have introduced the SiameseLSRTM framework to achieve highly accurate imaging. By employing two CNNs with shared weights, the Siamese network effectively extracts robust transformed representation from the observed and simulated data. This allows the loss function to accurately measure the differences between the two datasets, ultimately leading to enhanced imaging performance. To validate the effectiveness of SiameseLSRTM, we utilize two synthetic datasets and one field dataset from a land survey and test three objective functions—Euclidean loss, L1 loss, and L2 loss. The results indicate that irrespective of the selected objective function, SiameseLSRTM consistently yields more accurate imaging results than traditional LSRTM. Furthermore, the simulated data generated by the image obtained from the SiameseLSRTM aligns more closely with the observed data. From a computational perspective, SiameseLSRTM adds only a minimal overhead cost compared to traditional LSRTM. In conclusion, the proposed SiameseLSRTM serves as a valuable enhancement to the migration imaging method.

\section{Acknowledgments}
The authors sincerely appreciate the support from KAUST and the DeepWave Consortium sponsors. They also thank the SWAG group for fostering a collaborative research environment. The authors gratefully acknowledge the Supercomputing Laboratory at KAUST for providing the computational resources used in this work.

\bibliography{references.bib}
\bibliographystyle{unsrt} 






\end{document}